\documentclass[aps,pre,twocolumn,groupedaddress]{revtex4-1}

\usepackage{color}
\usepackage{amsmath}
\usepackage{graphicx}
\usepackage{subfigure}

\begin{document}

\title{Joint effect of ageing and multilayer structure prevents ordering in the voter model}

\author{Oriol Artime}
\email[]{oriol@ifisc.uib-csic.es}
\author{Juan Fern\'andez-Gracia}
\author{Jos\'e J. Ramasco}
\author{Maxi San Miguel}
\affiliation{Instituto de F\'isica Interdisciplinar y Sistemas Complejos IFISC (CSIC-UIB), Campus Universitat Illes Balears, 07122 Palma de Mallorca, Spain}

\begin{abstract}
The voter model rules are simple, with agents copying the state of a random neighbor, but they lead to non-trivial dynamics. Besides opinion processes, the model has also applications for catalysis and species competition. Inspired by the temporal inhomogeneities found in human interactions, one can introduce ageing in the agents: the probability to update decreases with the time elapsed since the last change. This modified dynamics induces an approach to consensus via coarsening in complex networks. Additionally, multilayer networks produce profound changes in the dynamics of models. In this work, we investigate how a multilayer structure affects the dynamics of an ageing voter model. The system is studied as a function of the fraction of nodes sharing states across layers (multiplexity parameter $ q $). We find that the dynamics of the system suffers a notable change at an intermediate value $ q^{*} $. Above it, the voter model always orders to an absorbing configuration. While, below, a fraction of the realizations falls into dynamical traps associated to a spontaneous symmetry breaking in which the majority opinion in the different layers takes opposite signs and that due to the ageing indefinitely delay the arrival at the absorbing state. 
\end{abstract}

\maketitle

\section{Introduction} 

Interaction times across social, biological and technical networks are commonly distributed in an irregular manner. In many of these systems, specially those for which the human or animal factor is in the equation, highly heterogeneous interaction patterns between basic components naturally emerge. Such heterogeneity is expressed in activation bursts, memory effects and long-tailed inter-event time distributions. \cite{barabasi2005origin,goh2008burstiness,radicchi2009human,karsai2011small}. The presence of non-Poissonian dynamics in models gives rise to a variety of interesting new phenomena like, for example, major changes in the speed of information diffusion and the behavior of spreading processes   \cite{iribarren2009impact,rosvall2014memory,boguna2014simulating, artime2017dynamics}. 
A good deal of attention has been devoted to temporal heterogeneities in monolayers, but we lack systematic studies in multilayer networks. To fill this gap, we study the voter model \cite{holley1975ergodic}, a paradigmatic example of a two-state out of equilibrium system with two equivalent absorbing states \cite{marro2005nonequilibrium}. It is one of the simplest models capturing random imitation and producing non trivial results\cite{suchecki2005voter}. It accounts for opinion competition \cite{castellano2009statistical,dornic2001critical,sanmiguel2005binary}, as well as for heterogeneous catalysis \cite{krapivsky1992kinetics,frachebourg1996exact} and species competition \cite{clifford1973model}. The model is sensitive to temporal inhomogeneities and to the update rule employed \cite{fernandez2011update, stark2008decelerating, takaguchi2011voter, perez2016competition,artime2017dynamics}. In particular, we focus here on a version of the voter model including node ageing that was recently introduced in Ref.  \cite{fernandez2011update}. The probability of an agent to update decreases as an inverse function of the time spent in a given state. The ageing has the role of an inertial force, making each node more conservative, i.e., less likely to change state as it has stayed more time in that certain state and it induces an ordering process in the system. In addition to the update rules, the interplay between topology and time heterogeneities is known to be crucial for determining the behavior of models in monolayer networks \cite{delvenne2015diffusion,artime2017dynamics}.

The effect of the coupling between topology and dynamics is further enhanced when the network structure is multilayered. Multilayer or multiplex descriptions of networks add a new degree of freedom to the topological properties and they are useful mathematical frameworks to describe scenarios where interactions occur in several contexts or different processes happen to the same set of nodes \cite{kivela2014multilayer, de2013mathematical, boccaletti2014structure}. In social networks, this, for instance, comprehends situations in which the relations can be of different nature (friends, family, coworkers, etc). In technological ones, it can correspond to connections with different bandwidth in each layer. The interactions on each of these contexts or layers rely on diverse network topologies, connecting a similar set of nodes and possessing a certain overlap in the links across layers. Note that a fraction of the nodes, but not necessarily all, can be present simultaneously in the several layers, bridging the information through them and playing a relevant role on the dynamics on these networks \cite{diakonova2014absorbing}. Beyond social systems, a similar framework can be used for temporal networks, with each layer corresponding to a time period \cite{holme2013temporal}, or spatial networks in which each node is a given area and the connections can represent different transport media \cite{barthelemy2011spatial}. The multilayer approach requires the introduction of new, and a generalization of already existing, concepts and metrics. Likewise, some of the features of dynamical models running on multilayer networks must be revised since this extra degree of freedom may yield new phenomenology. When studying processes in multilayer networks, a fair question is whether adding the information of the layers is necessary and will lead to results fundamentally different from a representation in a single network that aggregates all the links of the different layers. Recent works have addressed this question, confirming that multilayer structures and multiplexity give rise to phenomena that are not appreciated in one-layer networks in areas like epidemic spreading \cite{granell2013dynamical}, game theory and agent-based models \cite{cozzo2012stability,gomez2012evolution,gomez2012evolutionary,lugo2015learning, battiston2016robust} or transportation systems \cite{deDomenico2014navigability,gallotti2016lost}. Besides finding new results, it has also been shown that reducing the dynamics on a multiplex network into that on an effective single-layer network is not possible in certain cases as happens for the voter model \cite{Diakonova2016Irreducibility}.

In this work, we study how the presence of multilayer structure and realistic inter-event times in the form of ageing can affect the voter model dynamics. Indeed, these features bring new phenomena through a process of spontaneous symmetry breaking across layers. Our results show that the breaking of symmetry generates global states that trap the model dynamics and avoid the arrival at global consensus. The analysis is performed as a function of the fraction of nodes present simultaneously in different layers (multiplexity parameter) $ q $. Even though the model dynamics is local, nodes connected between layers update their state solidarily. Only the dynamics in the extreme cases $q = 0$, with no coupling between layers, or  $q = 1$, fully-coupling between layers (multiplex), is reproducible in the monolayer scenario. As the fraction of nodes with an intra-layer connection $q$ is changed, the voter model undergoes a modification of the dynamics. There is a specific value of $q$, $q^*$, below which the model has a finite probability to fall into a dynamically trapped configuration, while above $q^*$ the system always orders by coarsening towards a global consensus.

\section{Methods}

\begin{figure}
\centering
\includegraphics[width=6cm]{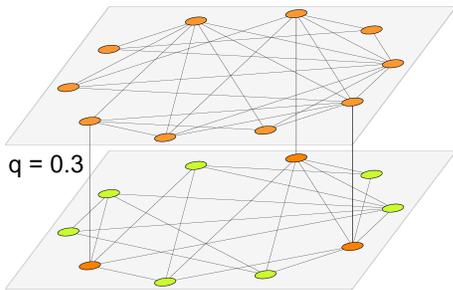}
\caption{Sketch of the two-layer topology used in the simulations. In each layer, we have a network with $ N $ nodes and a fraction of $ q $ nodes connected inter-layers. } \label{fig:3d-MultiLayer}
\end{figure}

\subsection{The model} 

The voter model is formulated in terms of a spin-like system, composed by agents endowed with a binary state (opinion) variable, say $ \pm 1 $. It is a paradigmatic model of opinion competition, which has its own universality class and it is solvable in many circumstances. \cite{dornic2001critical,krapivsky2010kinetic,barrat2008dynamical,slanina2003analytical,vazquez2008analytical}.
It is assumed that agents lack self-confidence, so they update their state by a process of random imitation of one of their neighbors' state in the interaction network. To be specific, the evolution in the standard voter model is made through the so-called random asynchronous update (RAU from now on). A basic step of the dynamics consists in picking a random agent and force her to adopt the state of a randomly chosen neighbor. The model has two equivalent absorbing configurations called consensus, which correspond to all nodes sharing the same state. On a monolayer, the way these configurations are reached has been widely studied and found to be crucially dependent on the network dimensionality, $ d $ \cite{krapivsky2010kinetic, suchecki2005voter, sood2005voter}. Thus, for $ d \le 2 $ the system orders, showing a coarsening process that drives the system to consensus, regardless of system size. Nevertheless, for dimensions $ d > 2 $, there is no coarsening process, but the system evolves to a metastable state and remains there until a finite-size fluctuation drags it towards an absorbing configuration.

\begin{figure*}
\centering
\includegraphics[width=14cm]{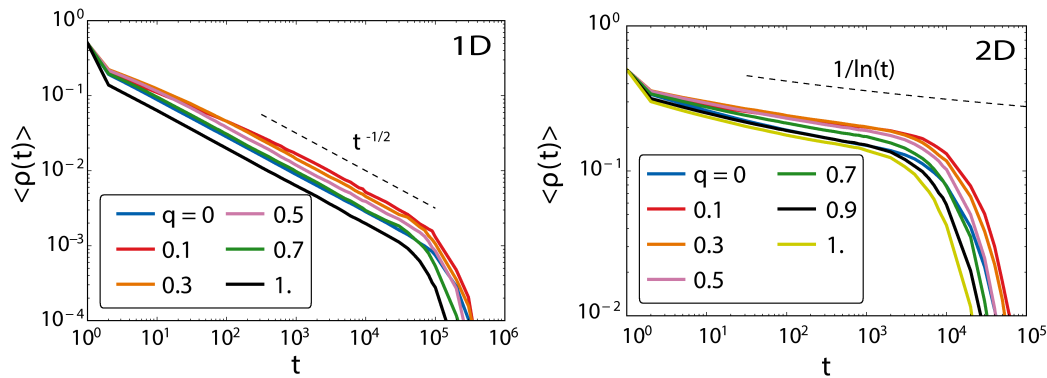}
\caption{Temporal evolution of the average fraction of active links for different values of $ q $. Left: One dimensional lattice of size $ N = 1000 $. Right: Two dimensional lattice of size $ N = 100^2 $. In both cases, the averages are taken over $ 1000 $ realizations.} 
\label{fig:Lattices}
\end{figure*}

In addition to RAU, we have implemented the so-called endogenous update for the voter model \cite{fernandez2011update}. In this case, each agent $ i $ keeps track of an internal clock measuring the persistence time $ \tau_{i} $, which stands for the time elapsed since her last change of state. All these times are initially set equal to $ \tau_{i} = 1 + b $, where $ b \geq 1 $ is a constant. Then each agent $ i $ undergoes an activation attempt at every time step. Such attempt occurs with probability $ p_i(t) = b/\tau_{i} $, otherwise nothing happens and  $ \tau_{i} $ increases in one unit. In case of success of the activation attempt, the agent selects one random neighbor and copies her state. There are two possibilities: either the node copies an agent with contrary state, changes her own opinion and thus resets her persistence time to the initial value, or she copies an agent with the same state so she does not change opinion and one unit of time is added to $ \tau_{i} $. The main non-trivial consequence of this update is the appearance of a coarsening process in high-dimensional networks, which drives the system towards consensus in contrast to what occurs with RAU. Indeed, the average of the fraction of active links ($\langle \rho \rangle$, links joining nodes with different states) and the inter-event time complementary cumulative distribution of the state changes ($C(\Delta t)$) behave as $ \langle \rho(t) \rangle \sim t^{-\alpha} $ and $ C(\Delta t) \sim \Delta t^{-\beta} $. Here $ \langle \cdot \rangle $ stands for average over realizations. For fully-connected networks $ \alpha = \beta = b $ and for complex networks (Erd\H{o}s-R{\'e}nyi, scale-free networks) $ 0 < \alpha \simeq \beta < b $. Note that as $ b $ increases, the tail of these quantities becomes more and more steep, resembling the RAU behavior for large enough values.

Along this work, we keep low values of $ b $. Specifically, we consider a natural extension of this update rule in a two-layer undirected network. Each layer has an activation probability for node $i$  $ p_{i}^l = b_{l}/\tau_i $, where $ l = {1,2} $ is the layer label and the constant $b_l$ can be different between the layers. The update is identical to the monolayer case, but now at every Monte Carlo step (MCS) the nodes undergo updates attempts in each layer. The order of the layers to update is chosen at random in each MCS. We define $ q $ as the intra-layer connectivity, which corresponds to the fraction of nodes that share state in both layers (see Figure \ref{fig:3d-MultiLayer}). This means that, if $ N $ is the number of nodes per layer ($ N_{1} = N_{2} \equiv N $), we have $ q\, N  $ \textit{common nodes}, which are forced to have the same state in both layers, and thus the same persistence time. We follow the evolution of the system by looking at the mean value between the two average fraction of active links of each layer, i.e., $ \langle \rho \rangle = \left( \langle \rho_{1} \rangle + \langle \rho_{2} \rangle \right)/2 $.

\subsection{RAU dynamics} 

Before exploring with detail the combined effects of multiplexity and ageing in the voter model, it is necessary a summary of results of the standard voter model in layered structures. We will use these results as a baseline for comparison. The case of two RAU voter models in complex networks embedded in multilayers is exhaustively studied by Diakonova and coathors in reference \cite{Diakonova2016Irreducibility}. In this work, two control parameters were used: the fraction of links that are simultaneously present in both layers $ \omega $ (overlapping parameter) and the degree of multiplexity $ q $, already introduced before. Similarly to the RAU voter model in monolayers, the system settles in a metastable state until a finite-size fluctuation forces it to reach consensus. One of their main results is that increasing $ q $ the average characteristic time scales to the the absorbing states decreases, while little effect is detected if the overlapping parameter changes. 
To complete the picture, we check with numerical simulations how the system behaves when endowed with RAU in each layer, but using low dimensional lattices. Indeed, low dimensionalities with the RAU voter model in single layers are of special interest, because the system orders even in the infinite size limit (coarsening process) for dimensions $ d \leq 2 $ and it is analytically solvable. If the RAU voter model is run on two lattices creating a duplex structure, does the low dimensional coarsening process breaks in this new topology? The way in which the two layers are connected is relevant, since nodes are located in space and there is a metric distance defined in the lattices. To avoid the redundancy of proximity effects, we choose the interlayer connection for each node at random. The average fraction of active links for multilayer lattices of dimension $ d = 1 $ and $ 2 $ as a function of the multiplexity parameter $q$ is plotted in Figure \ref{fig:Lattices}. Regardless of the dimension and the value of $ q $, the system orders by coarsening following the same functional form as in the monolayer lattices.

\section{Results}
\subsection{Complete graphs with ageing}

\begin{figure}
\centering
\includegraphics[width=8.6cm]{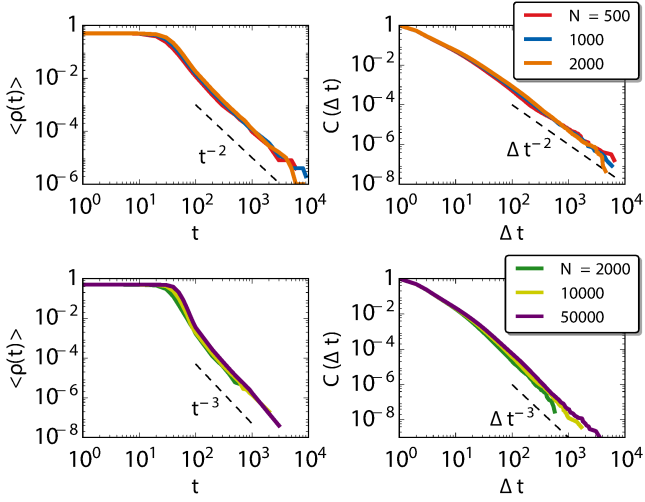}
\caption{Average fraction of active links as a function of time $\langle \rho (t) \rangle $ and cumulative inter-event time distribution $C(\Delta t)$ for a fully connected network with $ q = 1 $ and for different values of the activation coefficients. At the top $ b_{1} = b_{2} = 1 $. At the bottom $ b_{1} = 1 $ and $ b_{2} = 2 $. The tails of both functions decay as power-laws with exponent $ b_{1} + b_{2} $. The curves are averaged over $ 1000 $ realizations.} \label{fig:q1_FC}
\end{figure}

We consider first the simplest case with fully connected networks in the two layers because this will allow us to gain some analytical insights on the model behavior. Will running the endogenous update in the multilayer structure result in the same behavior than in the monoplex, or will it display new phenomenology?
The case with $ q = 0 $ does not require a detailed explanation: this is actually the model running on two independent monolayer networks. The fraction of active links as a function of time and the cumulative distribution of inter-event times, where the events are the change of state, display power-law decays with the same exponents as in monolayer networks. On the other extreme, the case $ q = 1 $, or multiplex structure, is more interesting. Numerically, we observe in Figure \ref{fig:q1_FC} that the average fraction of active links $\langle \rho (t) \rangle $ and the cumulative distribution of the inter-event times $C(\Delta t)$ for state changes decay as power laws with an exponent corresponding to  $ b_{1} + b_{2} $. The dynamics of the multiplex network is naturally accelerated respect to the one occurring in a monolayer network because the model in the multilayer duplicates the number of nodes updates. The decay with this particular exponent combination can be understood with the following analytical arguments.   
  
At each Monte Carlo time step, we try to update all nodes of a layer first and all nodes of the other layer after. The order of the update, which layer comes first, is chosen at random. Now that $ q = 1 $ all nodes are present in both layers, thus, in practice, each node tries to update twice per time step. Let us call $i$ to a generic node, and $i_1$ and $i_2$ to the node in layer $1$ and $2$, respectively, with identical spin value. Suppose that layer $ 1 $ is chosen first and layer $ 2 $ comes after. The possible update scenarios after a Monte Carlo step are: $(i_1,i_2)$, $(\overline{i_1},i_2)$, $(i_1,\overline{i_2})$, and $(\overline{i_1},\overline{i_2})$, where the name of the node in the duplets means that the node was selected to activate in the particular layer and the overlined name that it was not. Note that being selected to activate does not imply that the spin changed, this depends on the state of the neighbors. Recalling that the activation probability for node $i$ in layer $l$ goes as $p_i^l$ assuming that the order update in the layers was $1$ followed by $2$, one can write the probabilities of these four scenarios as
\begin{align}
\label{probconf}  
  P(i_1, i_2)    & = p_i^{1} \,\left(  s_i(t)\, p^{2,reset} + (1 - s_i(t)) \, p_i^{2} \right) , \nonumber \\
  P(i_1, \overline{i_2}) & = p_i^1 \, \left( s_i(t)\, (1 - p^{2, reset}) + (1 - s_i(t)) (1-p_i^{2}) \right) , \nonumber \\
  P(\overline{i_1}, i_2)    & = (1-p_i^1)\, p_i^2 ,   \\
  P(\overline{i_1}, \overline{i_2})    & = (1-p_i^1) \, (1-p_i^2),  \nonumber 
\end{align}
where $p^{l,reset}$ is the activation probability after a time reset in the contrary layer to $l$ and $s_i(t)$ stands for the fraction of neighbors with opposite opinion to $i$. It is worth recalling that the nodes have equal state in both layers, and thus $s_i(t)$ does not depend on the layer, but it does on time.

The effective probability of activation for node $i$ in a MC time step is then given by the sum of probabilities of the first three processes described in \eqref{probconf}, which yields
\begin{align}
\label{peff}
P_{act} & =  P(i_1, i_2) + P(i_1, \overline{i_2}) + P(\overline{i_1}, i_2) 
= 1 -  P(\overline{i_1}, \overline{i_2}) \nonumber \\
& = p_i^{1} + p_i^{2} - p_i^{1}\, p_i^{2} = \frac{b_{1}+b_{2}}{\tau_i} - \frac{b_{1}b_{2}}{\tau_i^{2}} .
\end{align}
If we consider the stochasticity in the layer choice, we should compute $ P_{act} $ assuming that layer 2 activates first, followed by layer 1. Still equation \eqref{peff} is symmetric to the interchange of layer labels and, therefore, we obtain the very same expression for $ P_{act} $. As can be seen in equation \eqref{peff}, $ P_{act} $ is always positive in the $ \tau $ domain of interest. Indeed, to satisfy that $ P_{act} (\tau) > 0 $, we need
\begin{align}
\tau (b_{1} + b_{2}) - b_{1}b_{2} \geq \left(\min\left(b_{1}, b_{2}\right) + 1\right) (b_{1} + b_{2}) - b_{1}b_{2} > 0, \label{ineq}
\end{align}
condition which is trivially satisfied. Equation \eqref{peff} tells us that for low values of the persistence time, the term with order in $ 1/\tau^2 $ dominates, while for large values of the persistence time, the term going as $ 1/\tau $ is the dominating one. In analogy to the monolayer, in which the decay exponent of quantities of interest is the coefficient of the activation probability, now in the multilayer we expect a decay exponent of $ b_{1} + b_{2} $ for long times, together with a correction for low times.

\begin{figure*}
\centering
\includegraphics[width=16cm]{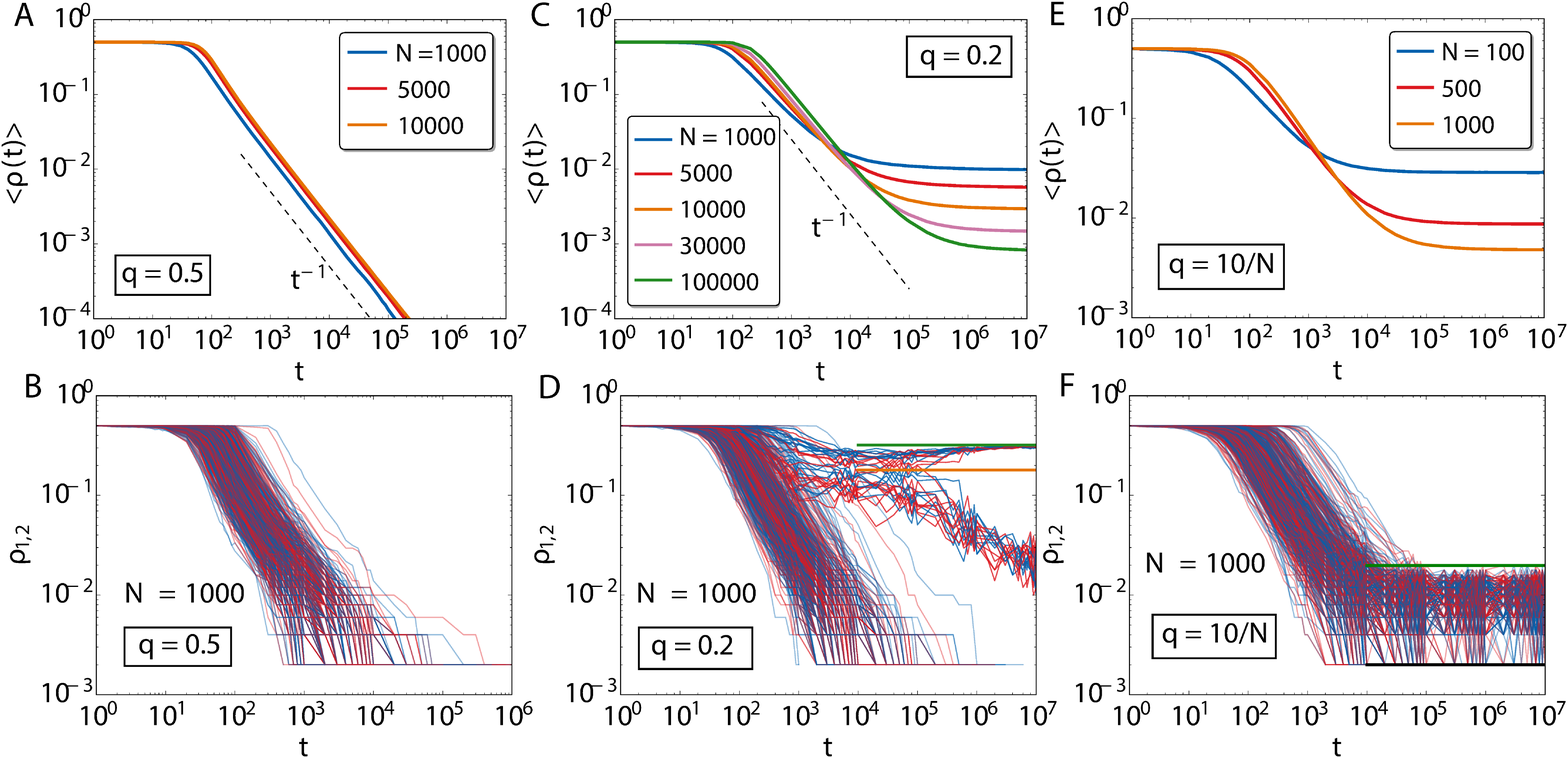} 
\caption{Average fraction of active links as a function of time over $1000$ realizations (panels A, C and E). In panels B, D and F, the corresponding values $ \rho_{1} $ (blue) and $ \rho_{2} $ (red) for each layer, for $300$ individual realizations of the dynamics. The fraction of active links is displayed for $ q = 0.5 $ (first column), $ q = 0.2 $ (second column) and $ q = 10/N $ (third column). The green horizontal line in panels D and F correspond to the predicted $\rho_{up}$ of equation \eqref{rhoup}. The orange line in panel D is the value of $\rho_{half}$ calculated in equation \eqref{rhohalf}. The black horizontal line in panel F corresponds to $\rho_{low}$ of equation \eqref{rholow}.}
\label{fig:Intermediateq}
\end{figure*}

When the parameter $q$ is not zero or one, we numerically find the three regimes depicted in Figure \ref{fig:Intermediateq}. For high values of $q$ ($q>q^{*}$), the model always orders by coarsening, while for lower values of $q$ ($ 0 < q \leq q^{*} $), plateaus are observed in $\langle \rho \rangle$ in the large time limit. The larger the system size is, the lower the plateau. To understand which is the origin of these plateaus, it is helpful to inspect individual realizations of the system dynamics (see videos included as Supplementary Information). The bottom row of Figure \ref{fig:Intermediateq} shows the order parameter $\rho_{1,2}$ in each of the layers for different realizations. We observe some realizations in which $\rho_{1,2}$ tend to zero as the system orders by coarsening, following a power law. There are others in which two branches emerge in a reciprocal way: the fraction of active links for one of the two layers, $\rho_{1(2)}$, remains almost constant, indicating the survival of a fraction of nodes with opposite opinion to the majority one. This goes along with the ordering by coarsening of the other layer as $\rho_{2(1)}$ tends to zero, following again a power law, with smaller exponent though. In practice, what is occurring here is an spontaneous symmetry breaking between the two layers: each layer orders towards different opinions, so the system ends up in a mixed state. It is then when common nodes come at play, slowly ordering by coarsening towards one of the two opinions, what we call the opinion of the dominant layer. This configuration is able to trap the system dynamics in the long time limit due to node ageing and to preventing the coarsening towards global consensus. We define $ q^{*} $ as the largest value of $q$ for which we observe realizations that do not approach global consensus by coarsening, i.e., realizations that trap the dynamics. Additionally, for very low values of the multiplexity parameter, in the regime $0 < q \le q_{0} (N) \sim \mathcal{O} (1/N) $, there is a finite size crossover, that disappears in the thermodynamic limit. In this regime of $ q $ values, the two layers arrive again to a fully-ordered state, each one to different states, except for the common nodes. These ones keep oscillating randomly from one consensus state to the other, so there is no coarsening process between them. 

It is possible to estimate the asymptotic values of the plateaus. The relevant nodes are the common ones to both layers, so in the large time limit we assume that one layer is ordered in one opinion and the other layer ordered in the other opinion. For the common nodes, we do not assign them a priori any state. In the regime $ 0 < q \leq q_{0} (N) $, we see that the trapped configurations bounce between two values, $ \rho_{low} $ and $ \rho_{up} $ (horizontal green and black lines in Figure \ref{fig:Intermediateq}F). We assume that one common node holds a state and the rest of common nodes are in the contrary state. The value of the plateau in the fraction of active links of the dominant layer is given by
\begin{equation}
  \rho_{low} = \frac{(N-1)}{N\, (N-1)/2} = 2/N. \label{rholow}  
\end{equation}
This equation is just the ratio between the number of connections of the contrary solitary node with the rest of nodes and the total number of links in the complete graph. The other plateau, $\rho_{up}$, corresponds to the dominated layer and can be computed following the same logic. The value is given by
\begin{equation}
 \rho_{up} = \frac{q\, \left(N-1\right)\left[N-q\left(N-1\right)\right]}{N\left(N-1\right)/2} \sim 2\, q\, (1-q),\label{rhoup}
\end{equation}
where in the last step we performed the limit $ N \rightarrow \infty $. In the regime $ q_{0} (N) < q \leq q^{*} $, there is a third plateau at $\rho_{half}$, that occurs earlier in time, while the dominance of the layers is still to be defined. The fraction of active links remains around  $\rho_{half}$ for a time that is longer for lower values of $q$, and then it tends to $ \rho_{up} $ in the dominated layer and to $\rho_{low}$ (zero in the thermodynamic limit) for the dominant layer. This plateau  corresponds to the value in which half of the common nodes are in one opinion and the rest in the other, so we can write that
\begin{equation}
 \rho_{half} = \frac{\frac{q\, N}{2}\left(N-\frac{q\, N}{2}\right)}{N\left(N-1\right)/2} \sim q\left(1-\frac{q}{2}\right),\label{rhohalf}
\end{equation}
again taking the thermodynamic limit in the last step. The values predicted by these equations are shown in the individual curves of Figure \ref{fig:Intermediateq}, presenting a good agreement with the numerical simulations.

The appearance of the dynamically trapped configurations is a consequence of ageing and multilayer network structure. Both ingredients are needed simultaneously to develop the plateaus, as independently the ageing in the monoplex always leads to consensus through coarsening and the standard voter model with RAU in the multiplex reaches consensus by means of a finite size fluctuation, independently of $ q $. Due to ageing, nodes become older, so they are less likely to try to update their state as time increases. If we combine this with a small enough value of $ q $, the realizations in which there is an ordering of the layers towards different opinions get stuck in this situation and they are not able to reach the global consensus. At this stage, common nodes play the important role. They limit the maximum amount of order in the layers. Due to them, there are always $ q \,N $ states in every layer that are forced to be in agreement with the other layer. In terms of the total magnetization $ m(t) = N^{-1} \, \sum_{i} x_{i}(t) $ where $x_i$ is the opinion of node $i$, we have that $ |m| \sim 1$ ($ |m| \sim 1 - 2\, q$) for the dominating (dominated) layer at large times.

\begin{figure*}
\centering
\includegraphics[width=16cm]{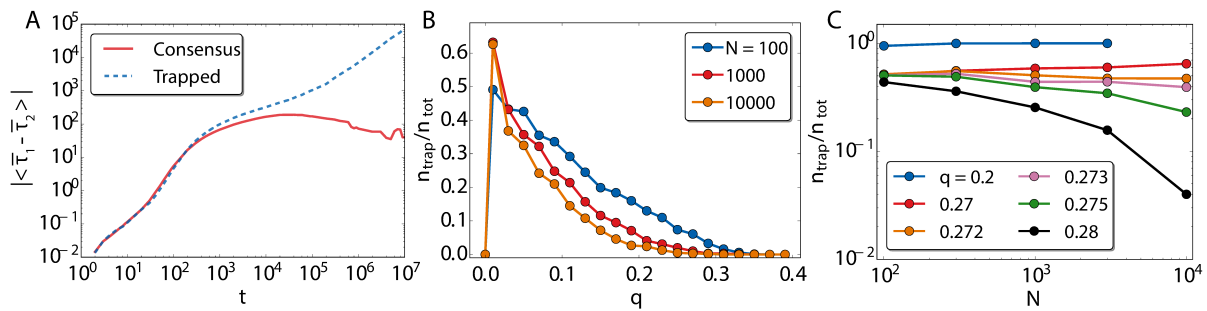} 
\caption{In A, difference of average persistence times as a function of time for those realizations that reach the global consensus state and those that remaining trapped, employing $N=1000$ and $q=0.1$ and averaging over $ 1000 $ realizations. In B,  the fraction of trapped configurations as a function of $ q $. Each point is computed over $1000$ realizations. Note the singularity of $ q = 0 $, in which there is no possibility of trapping since the layers are disconnected and their own evolution is always a coarsening process to consensus. In C, the fraction of trapped configurations as a function of $ N$ for several values of $q$. Each point calculated over $1000$ realizations.} \label{fig:Intermediateq2}
\end{figure*}

To quantify the role of ageing in the emergence of these dynamically trapped configurations, we introduce the mean persistence time $ \overline{\tau_{l}} = N^{-1} \sum_{i} \tau_{i, l} $, which gives us an idea of how old the nodes are in layer $ l $. Here, the overline stands for the average over one network realization. Thus, the quantity $ \langle \overline{\tau_{1}} - \overline{\tau_{2}} \rangle(t) $, where again the brackets mean the average over realizations, provides an estimation of how different the age of nodes are in both layers. We can see in Figure \ref{fig:Intermediateq2}A that, depending on whether the simulations reach the global ordered state by coarsening or remain trapped, the behavior in this quantity is very different. Initially, the mean persistence times of each layer are similar and they increase accordingly. At intermediate times, a change of behavior takes place, either $ \langle \overline{\tau_{1}} - \overline{\tau_{2}} \rangle(t) $ remains constant (global consensus by coarsening) or it keeps growing (trapped configurations). Thus, when the system falls in the dynamically trapped configuration, there is a imbalance between the age of two layers, which grows in time. The older nodes correspond to the dominant ordered layer $ \left( |m| \sim 1 \right) $. 

The fraction of realizations falling into the trapping state, $n_{trap}/n_{tot}$, can be also analyzed as a function of $q$. As can be seen in Figure \ref{fig:Intermediateq2}B, it decays as $q$ increases until vanishing for $q^*$. For the system sizes considered in our simulations, $q^*$ lies between $0.25$ and $0.35$. As the size increases $q^*$ gets smaller, even though it is still well over $0.2$. The shape of the curve of $n_{trap}/n_{tot}$ in function of $q$ seems to get stable for large $N$. In order to go further and to perform a finer system size analysis, it is necessary to select the initial conditions in such a way that the number of realizations falling in the trapping states is enhanced. Otherwise, the numerical effort to obtain an acceptable resolution for $n_{trap}/n_{tot}$ becomes prohibitive in the large $N$ limit. In all the figures so far, we have set the initial values for the node spins at random: $+1$ or $-1$ with probability one half. The trapping configurations take place when is ordered in one state while the other is ordered in the contrary state. We can recreate this by assigning $+1$ to all the nodes in one layer, $-1$ to all of the other and selecting one of the two values for all the common nodes that become thus incorporated in the dominant layer. This initial condition notably improves the number of realizations falling in the trapping states. With this tool, we explore the values of $q$ in Figure  \ref{fig:Intermediateq2}C for several system sizes. The objective is to determine whether there exist a value of $q^*$ below which the trapping states are stable in the thermodynamic limit and above which they become unstable with the system falling into the global consensus by means of coarsening. This is precisely what we find for a value of $q^*$ laying between $q = 0.27$ and $0.275$.

\begin{figure*}
\centering
\includegraphics[width=16cm]{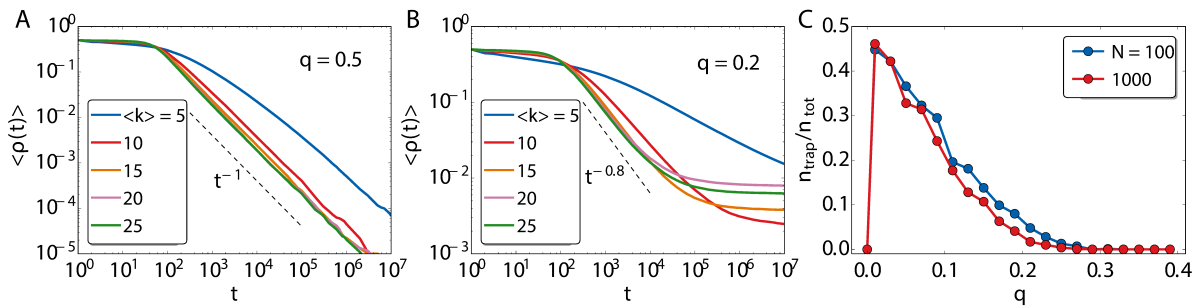}
\caption{Average fraction of active links as a function of time for different values of $ \langle k \rangle $ with $ q = 0.5 $  (A) and $ q = 0.2 $ (B). The system size is $ N = 1000 $ and the curves are averaged over $ 1000 $ simulations. In C, the fraction of trapped realizations $n_{trap}/n_{tot}$ starting from random initial conditions averaged over $1000$ realizations for two network sizes and with  $\langle k \rangle = 14$.}  \label{fig:ER}
\end{figure*}

\subsection{Poissonian and scale free random networks} 

So far we explored the behavior of the system in complete graphs. In this section, we check whether the results obtained before are robust under other topologies. Unlike in complete graphs, the nodes in complex networks may not be  equivalent because the local neighborhood is different for each node. To build the networks, we use a configurational model approach \cite{molloy1995critical} in which the nodes' degrees are assigned from random extractions of a given degree distribution and then connections are established between randomly chosen node pairs until satisfying all the degree constraints. A maximum degree $k_{max} = \sqrt{N}$ has been used to prevent the introduction of degree-degree correlations \cite{catanzaro2005}, and a minimum degree of at least $k_{min} = 2$ to ensure that the network is not fragmented. Two main degree distribution families are considered: a Poissonian distribution with average degree  $ \langle k \rangle$ and power-law decaying distributions $P(k) \sim k^{-\gamma}$. 

Given that the fluctuations in the degree are small, the configurational model with Poissonian degree distribution produces networks with statistically equivalent nodes. We have employed the same average degree in both layers: 
$ \langle k_1 \rangle = \langle k_2 \rangle = \langle k \rangle$ and randomly connected the fraction $q$ of nodes between the two layers. The multilayers with $ q = 0 $ and $ q = 1 $ do not bear surprises respect to the fully connected graphs. In both cases, the model goes to consensus by coarsening. In $q = 0$, $\langle \rho \rangle$ drops as a power-law even though the exponent is lower and the dynamics slower than in complete graphs. For multiplex networks $\left( q = 1 \right) $, the behavior of $\langle \rho \rangle$ is weakly dependent on $\langle k \rangle$ in the range of values that we have explored. For intermediate values of $q$ (Figure \ref{fig:ER}), the full picture described in the complete graph case repeats: There is a regime for high $q$ values, $ q > q^{*}$, in which the system orders towards global consensus in a similar ways as for $q = 1$ (see Figure \ref{fig:ER}A). We see that there is a little dependence on $ \langle k \rangle $: only when it is very low the dynamics slows down. By decreasing $ q $ we enter into a different regime $ \left( q < q^{*} \right) $ in which dynamically trapped configurations are possible (Figure \ref{fig:ER}B). In this range of values of $q$, the dependence on $ \langle k \rangle $ is stronger. By increasing the mean degree, we see that the dynamics becomes faster and the plateau heights tend to increase accordingly (except when we reach to very high mean degree). One can also depict a diagram with the fraction of trapped realizations as a function of $q$ (Figure \ref{fig:ER}C). The behavior is similar to the one observed for complete graphs: the existence of $q^*$ above which the system dynamics do not get trapped and the curves get stable as $N$ increases. In theory, a quantity that may be relevant when having multilayer networks is the overlapping parameter $ \omega $, which represents the fraction of links present in both layers. However, we have found that the results shown here do not change qualitatively for different values of $\omega$.

\begin{figure*}
\centering
\includegraphics[width=16cm]{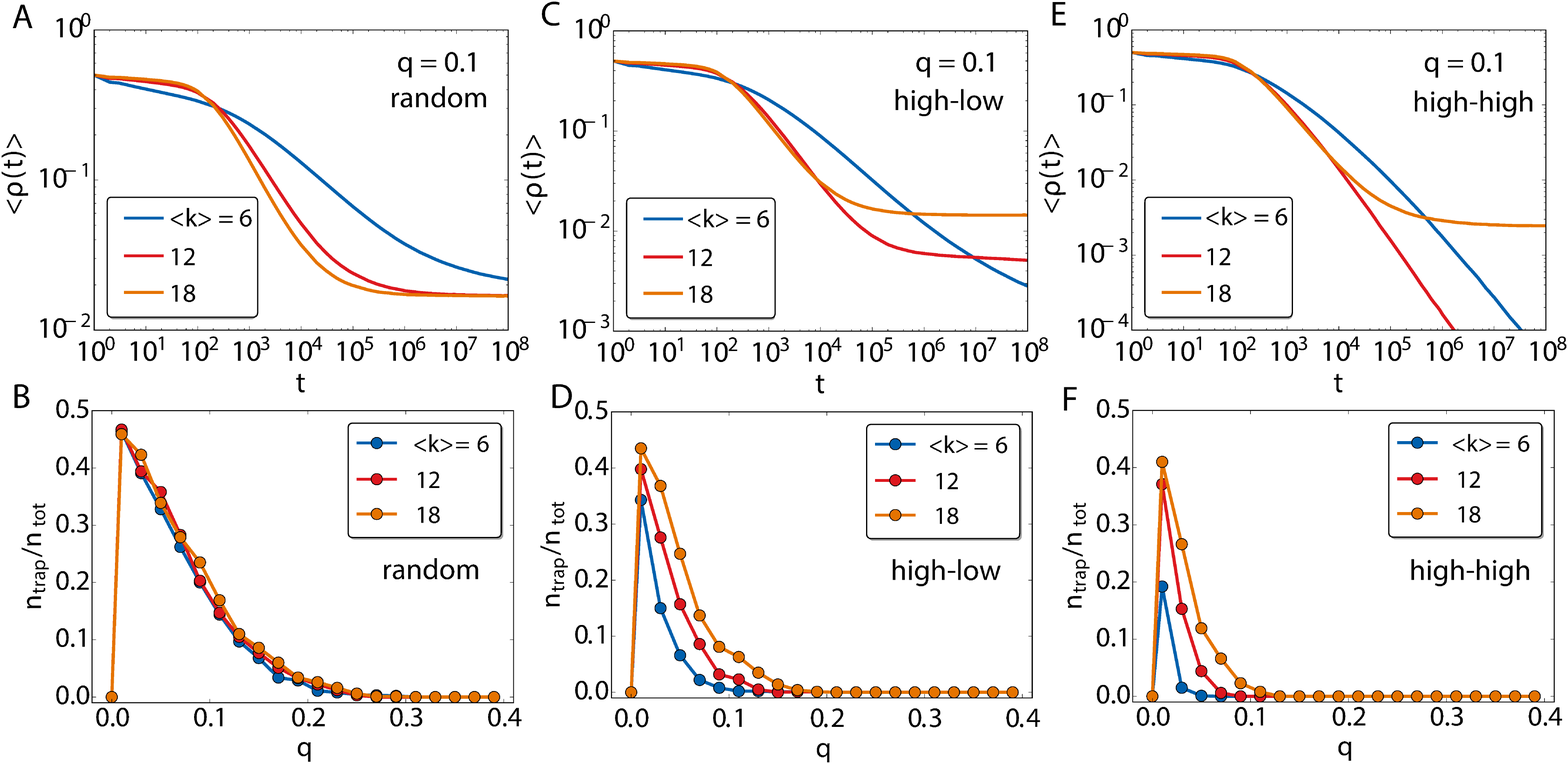}
\caption{In A, C and E, average fraction of active links as a function of time for different values of $ \langle k \rangle $ with $ q = 0.1 $. In A, the interlayer node connections are set at random, regardless of the node degrees, while in C the nodes are connected following a high-low degree order and in E following the opposite high-high order. In B, D and F, the corresponding diagrams with the fraction of realizations falling in the trapping states as a function of $q$. The network topology within each layer is scale free with degree distribution $P(k) \sim k^{-2.5}$. The network sizes are $N = 2000$ and $\rho$ is averaged over $1000$ realizations.} 
\label{fig:SF}
\end{figure*}

Homogeneous complex network topologies in each layer do not change the features found for the ageing voter model embedded in the multilayer, with complete graphs. The large differences in the node degrees open, however, new opportunities when building the multilayer network. To figure out if the heterogeneity in the connections have an impact, we display in Figure \ref{fig:SF} the average fraction of active links for three types of multilayer scale-free networks. The exponent of the degree distribution is set at $\gamma = -2.5$ within each layer and the subnetworks are drawn randomly from the configurational model approach. In particular, we consider three different interconnection procedures between layers. In the random procedure, the nodes belonging to the set of common nodes are chosen at random, irrespective of their degree. The high-low procedure selects the $ q\,N $ nodes with highest degree of a layer and connects them with the $ q\,N $ nodes with lowest degree of the other layer. Finally, the third procedure, which we call high-high, corresponds to connect the $ q\,N $ nodes of highest degree of each layer.

We show in Figure \ref{fig:SF}A and B the results for the random procedure. For $q = 0.1$, $\langle \rho (t) \rangle $ has a plateau for large times, so the trapping states for the dynamics exist. Furthermore, the diagram of the fraction of realizations falling in the trapping states in Figure \ref{fig:SF}B is similar to those obtained for fully connected networks in Figure  \ref{fig:Intermediateq2} and Poissonian random networks in Figure \ref{fig:ER}. Even the value of $q^*$ is maintained in a close range of values. The presence of heterogeneity in the node degrees does not seem thus to have a large impact in this case. However, this is no longer true when the interlayer links are correlated. Figures \ref{fig:SF}C-F show the simulation outcomes for these two connection procedures. The trapping states are still present in both cases, although the plateaus of $\langle \rho(t) \rangle $ for $q = 0.1$ appear at different levels: lower than random for high-low networks (Figure \ref{fig:SF}C) and even lower or not a at all for high-high ones (Figure \ref{fig:SF}E). To understand the effect of the correlated connection procedures and the mean degree, we display the fraction of trapped realizations diagrams in Figure \ref{fig:SF}D and \ref{fig:SF}F. In both cases, the full diagram is displaced to the left, toward lower values of $q$, as also is $q^*$. The reason for this difference respect to the random multilayer configurations is in the role played by the hubs. While in the random multilayers, especially for low values of $q$, the hubs are unlikely selected to have an inter-layer connection, in the case of the correlated multilayer networks they are always selected either to connect to other hubs or to low degree nodes. The hubs in the voter model are the most influential nodes, since their opinion has a larger chance to be copied by the rest of the layer. The formation of the trapping states passes through having one layer in one majority opinion and the other in the opposite one. If the hubs are involved in the inter-layer connections, this configuration is harder to reach. This effect naturally displaces $q^*$ toward lower values, so much so if the correlations are of the type high-high since getting both layers to disagree is more difficult.

\subsection{Mixed update rules}

One of the physical motivations for moving to multiplex topologies is to include the possibility of having different rhythms of interactions in every layer, or that the underlying mechanism for the evolution of information can be different. Interesting results have been obtained in for example in Refs. \cite{granell2013dynamical,czaplicka2016competition,vilone2012social,vilone2014social}. To address this kind of question, we study next the scenario in which there are two different updates in the multiplex: the RAU and the endogenous update. Does the system show a coarsening process? How does the number of common nodes affect the timescales of the model in this setup? In case of competing outcomes of the dynamics for the different updates, what is the result of this competition?

For the sake of simplicity, we study a complete graph per layer. In this case, for a single layer, we have opposite results depending on the update rule employed. On one hand, for RAU the system does not order in the thermodynamic limit, but stays in a disordered metastable state until a finite size fluctuation brings it to one of the two consensus configurations in a characteristic time of the order of the size of the system. On the other hand for the endogenous update there is a rather slow coarsening process. 

\begin{figure}
\centering
\includegraphics[width=8.6cm]{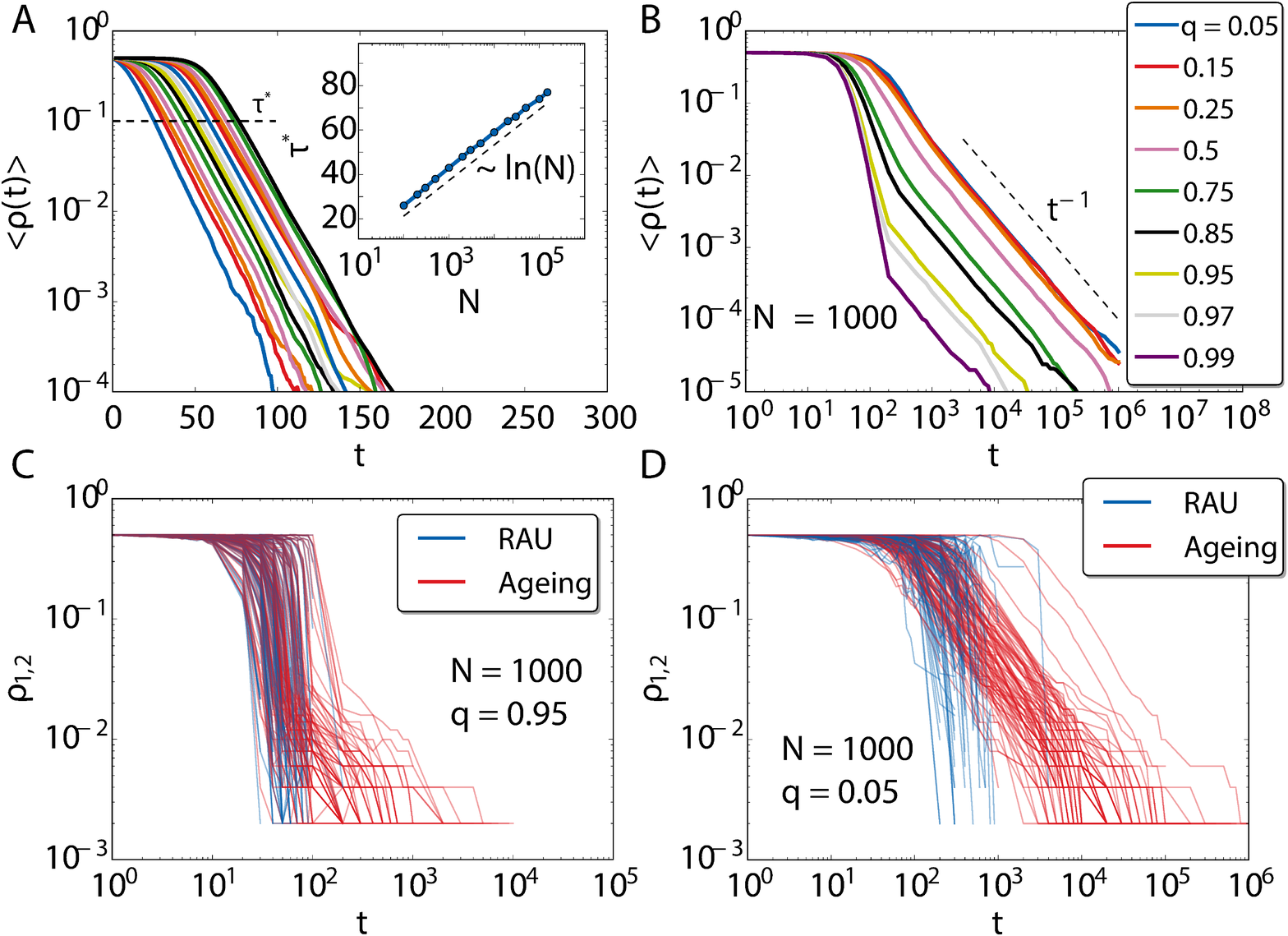}
\caption{Mixed updates on two complete graphs forming a partial multiplex, one layer with endogenous update ($b=1$) and the other with RAU. In A, average fraction of active links as a function of time for different system sizes $ N $, for $ q = 1 $. In the inset, the time to cross $ \tau^{*} = 10^{-1} $ as a function of $ N $. In B, average fraction of active links as a function of time for different values of $ q $; the system size is $ N = 1000 $.  Averages over $ 1000 $ simulations. In C and D, $ 2\times100 $ individual trajectories for $ q = 0.05 $ and $ q = 0.95 $, respectively. The density of active links is separated by layer, showing in red the one with ageing and in blue is the one with RAU. } \label{fig:endvsrau}
\end{figure}

In Figure \ref{fig:endvsrau} we display the results of the voter model with endogenous update ($b=1$) in one layer and RAU in the other one. On panel A, we show the behavior of the system when all nodes are shared between layers, $ q = 1 $. As we see, the order parameter decays exponentially, resembling the case of RAU in a monolayer, but now with a characteristic time that does not depend on system size. Still there is system size dependence: the exponential decay appears after a time $\tau$ that scales as the logarithm of system size $N$, i.e., very slowly. We evaluated this by measuring the time $\tau^*$ that takes the system to reach the reference value $\langle \rho \rangle =0.1 $ (inset of Fig. \ref{fig:endvsrau}A). In this case, and in cases where $q$ is close to one, both layers are strongly coupled and the fastest layer to reach an absorbing configuration (the one with RAU) dictates the global dynamics. This can be seen in Fig. \ref{fig:endvsrau}C, in the evolution of $\rho$ for both layers for single realizations for a large value of $q$ (blue for the RAU layer and red for the endogenous update layer). In Figure \ref{fig:endvsrau}B, we show the global average fraction of active links as function of time for several values of $ q $ and for a fixed $ N=1000 $. While $ q $ is close to $ 0 $, it decays as a power law of exponent $ -1 $. Since in this case both layers are weakly coupled, the RAU layer orders exponentially fast and all the contribution to $ \langle \rho_{m} \rangle $ comes from the endogenous layer, which keeps active for a longer time and orders with its known coarsening process in a monolayer. This result is further reinforced by Fig. \ref{fig:endvsrau}C where we see individual realizations for a small value of $q$ where the density of active links is plotted separately for both layers (RAU layer in blue and endogenous update layer in red). As $ q $ increases, the layers interact more and more and $ \langle \rho_{m} \rangle $ reproduces a mixture of the behaviors of the RAU and the endogenous update. Initially it decays exponentially until by the power law decay takes over. In these cases, the RAU dominates initially, imposing a consensus configuration in its time scale for the nodes of its layer and the common nodes. Then, the system needs a longer timescale to reach full consensus, following the typical scaling law of the endogenous update for those few not shared nodes of the endogenous layer.

\section{Discussion} 

The standard random asynchronous update (RAU) of the voter model has been widely studied along past years in different topologies, like lattices, complex networks, complete graphs or multilayer structures. However, there is still missing a complete understanding of the effect of real interaction patterns and their interplay with the topology. Regarding the temporal dimension, in RAU it is assumed that nodes are memoryless, so they update their state on average once per time step. Real systems, nevertheless, present strong temporal heterogeneities and non Markovian approaches must be considered in the modeling framework. There are several ways to generalize in order to include this phenomenon, one of which is the ageing in the agents. This translates into a probability of activation that falls according to the time spent in the same opinion, and it drastically changes the behavior of the model. Ageing in the voter model has been previously studied in the case of complex topologies, where the system orders towards consensus following a power law in its order parameter (fraction of active links). This is in contrast to the RAU voter model, which reaches the absorbing state by the effect of finite size fluctuations. Regarding the topological part, in last years multilayered networks has been proven as an important tool to model scenarios in which processes of different nature run in the same system or different kind of updates play a role in the evolution of the dynamics. This compresses, for instance, the modeling of systems with different types of interactions in social and technological networks. Dynamical processes running on top of these networks produce new phenomena that has been unobserved so far.

In this work, we have studied non-standard updates of the voter model in a two layered network, with fully-connected or complex topologies embedded in each layer. We performed our analyses in function of the degree of multiplexity $ q $, which is the fraction of nodes forced to have the same state in both layers. The RAU voter model in a two layered complex network always reaches consensus, with some influence of $ q $ (and the overlapping parameter). When, on the contrary, one layer is provided with ageing and the other stays memoryless, there may be a competition of time scales in the system. We found that the global consensus is again the final fate of the system. The combined evolution of the system is determined by the multiplexity parameter: for low $ q $, the ordering of the layer with endogenous update dictates the behavior, in which the system spends most of its lifetime ordering by coarsening. On the opposite, for large $ q $, the RAU layer dominates and orders the majority of the global system by finite size fluctuations, which results in a much faster time scale due to the exponential evolution of the order parameter. However, a much richer and intriguing scenario appears when updates with ageing are considered in the two layers. In this case, depending on the value of $ q $ the system does not necessarily orders. There exist states generated by spontaneous symmetry breaking that are able to trap the dynamics. Such states would be metastable in the RAU dynamics but here thanks to the combination of ageing and the multilayer topology they become stable, being reinforced as time passes. The trapping states appear below a certain value $ q^{*} $, in which there is a finite probability of the system to get trapped. When this occurs, the order parameter of one layer tends to a finite plateau value that we have computed analytically, while the order parameter of the other layer tends to vanish in the thermodynamic limit following a coarsening process. We found that our results are robust against changes in the network topology. In the case of complete graphs in the layers, the dynamics is faster and the phenomena are easier to appreciate than in random networks.

\section{Acknowledgements}

Partial financial support has been received from the Spanish Ministry
of Economy (MINECO) and FEDER (EU) under the project ESOTECOS
(FIS2015-63628-C2-2-R).

\bibliography{ArXiv_multiplex_v2}

\end{document}